\documentclass[graybox]{svmult}

\usepackage[utf8]{inputenc}
\usepackage[T1]{fontenc}

\usepackage{mathptmx}       
\usepackage{helvet}         
\usepackage{courier}        
\usepackage{type1cm}        
\usepackage{makeidx}         
\usepackage{graphicx}        
\usepackage{multicol}        
\usepackage[bottom]{footmisc}

\usepackage{amsmath}
\usepackage{amsfonts}
\usepackage{amssymb}
\usepackage{bm}
\usepackage{multirow}
\usepackage{diagbox}
\usepackage[textsize=footnotesize]{todonotes}

\makeindex             

\definecolor{darkblue}{rgb}{0,0,0.6}
\definecolor{darkred}{rgb}{0.6,0,0}
\usepackage[colorlinks=true,urlcolor=darkblue, citecolor=darkblue,linkcolor=darkred,hyperfootnotes=false]{hyperref}

\newcommand{\be}{\begin{equation}}
\newcommand{\ee}{\end{equation}}
\newcommand{\ba}{\begin{eqnarray}}
\newcommand{\ea}{\end{eqnarray}}
\newcommand{\bit}{\begin{itemize}}
\newcommand{\eit}{\end{itemize}}
\newcommand{\Ha}{\mathcal H}

\newcommand{\R}{\mathbf{R}}

\newcommand{\fv}{\mathbf{f}}
\newcommand{\kv}{\mathbf{k}}
\newcommand{\rv}{\mathbf{r}}
\newcommand{\uv}{\mathbf{u}}
\newcommand{\vv}{\mathbf{v}}

\newcommand{\ind}[1]{_{\mathrm{#1}}}

\begin{document}

\title*{Mechanical factors affecting the mobility of membrane proteins}
\author{Vincent Démery and David Lacoste}
\institute{Vincent Démery \at Gulliver, ESPCI Paris, PSL University, CNRS, 75005 Paris, France, \email{vincent.demery@espci.psl.eu}
\and David Lacoste \at Gulliver, ESPCI Paris, PSL University, CNRS, 75005 Paris, France \email{david.lacoste@espci.psl.eu}}
%
%
\maketitle

\abstract*{}

\abstract{
The mobility of membrane proteins controls many biological functions.
The application of the model of Saffman and Delbrück to the diffusion of membrane proteins does not account for all the experimental measurements.
These discrepancies have triggered a lot of studies on the role of the mechanical factors in the mobility.
After a short review of the Saffman and Delbrück model and of some key experiments, we explore the various ways to incorporate the effects of the different mechanical factors.
Our approach focuses on the coupling of the protein to the membrane, which is the central element in the modelling.
We present a general, polaron-like model, its recent application to the mobility of a curvature sensitive protein, and its various extensions to other couplings that may be relevant in future experiments.
}

\section{Introduction}

Cell membranes are barriers that separate the cytoplasm from the external world. Through compartmentalization, they allow highly selective 
biochemical reactions to take place in their internal volume which would essentially never occur in the absence of such barriers. 
Far from being inert, biological membranes thus play a key role in many functions such as signaling, cell division, 
or energy production in cellular organelles. 

Many of these biological functions involve membrane proteins, which form a vast family of essential proteins. 
While some of them are embedded permanently in the membrane,
 others transiently bind to it in order to perform a specific task and then unbind from it when the task is done. 
When they are inserted in the membrane, membrane proteins typically diffuse laterally in the fluid environment of the lipid 
membrane. 
This lateral diffusion is an essential aspect to their function, in the frequent case that membrane proteins must 
interact or form clusters with other membrane proteins. Membrane proteins typically diffuse in a crowded environment of other lipids and proteins. 
Modeling the various factors (mechanical or biochemical) affecting the mobility of membrane proteins is  
a challenging issue, but one that should be addressed in order to 
properly understand their biological function.

In this chapter, we review some of the experiments and theories, which have been devoted to the mobility of membrane proteins. 
In the next section \ref{sec:SD}, we present the pioneering work of P. G. Saffman and M. Delbr\"{u}ck (SD) on the mobility of membrane 
proteins, which is followed in section \ref{sec:Exp-SD} by a discussion of the experiments which have either tested the model or pointed out its limitations. 
In section \ref{sec:Coupling}, we investigate the crucial couplings between the membrane protein and the membrane. Then, in section \ref{sec:Theory} 
we present some of the main theoretical ideas or models which 
have been put forward to understand the mobility of membrane proteins beyond the SD model. 
We end up with a discussion on future perspectives.

\section{The Saffman and Delbr\"{u}ck hydrodynamic model (SD)} \label{sec:SD}

Brownian motion plays an essential role in biological processes. Since the work of Einstein~\cite{Einstein1905} and the pioneering experiments of Perrin~\cite{Perrin1909}, the observation of diffusing objects 
has emerged as a mean to extract the rheological properties of the surrounding medium or the probe particle size. The theoretical investigation of diffusion of 
proteins within membranes has been studied widely going back to P. G. Saffman and M. Delbr\"{u}ck (SD).  They investigated the hydrodynamic drag acting 
on a membrane inclusion of radius $a_\mathrm{p}$ moving in a membrane described as a two dimensional fluid sheet of viscosity $\mu_\mathrm{m}$; which is itself in contact with  a less viscous fluid of viscosity $\eta$ \cite{Saffman1975}. The two dimensional surface viscosity of the membrane $\mu_\mathrm{m}$
is the product of the membrane thickness $h$ by its three dimensional viscosity $\eta_\mathrm{m}$, $\mu_\mathrm{m}=h \eta_\mathrm{m}$.  
The velocity field inside the membrane is exactly two-dimensional 
but it is hydrodynamically coupled to the external fluid. 
Using singular perturbation techniques, which are valid when $\eta_\mathrm{m} h \gg \eta a_\mathrm{p}$, Saffman and Delbr\"{u}ck (SD) 
obtained the diffusion coefficient $D_{0}$ 
for the translational Brownian motion of a cylindrical inclusion of radius $a_\mathrm{p}$ in the membrane:
\be
D_{0} = \frac{k_B T}{4\pi \mu_\mathrm{m}} \left[\log\left(\frac{\ell}{a_\mathrm{p}}\right) - \gamma \right],
\label{SD-D}
\ee 
where $k_B T$ is the thermal energy, $\gamma$ is Euler's constant and $\ell = \mu_\mathrm{m}/\eta$ is the SD length.
The expression (\ref{SD-D}) corresponds to the choice of a no-slip boundary condition at the surface of the inclusion; for the 
alternate choice of zero tangential stress boundary condition, a factor $1/2$ should be added inside the bracket.
Saffman and Delbr\"{u}ck also derived an expression for the rotational diffusion coefficient, which unlike the translational diffusion coefficient only depends on the viscosity of the membrane: $D_R=k_B T/4 \pi h \mu_\mathrm{m} a_\mathrm{p}^2$.

The SD dimensionless parameter $\epsilon = \ell/a_\mathrm{p}$ represents physically the ratio of the hydrodynamic resistances of 
the inclusion in the fluid membrane and in the external fluid. Indeed the former is of the order of 
the lateral area of the inclusion $2 \pi a_\mathrm{p} h$ times the membrane shear stress $v \eta_\mathrm{m}/a_\mathrm{p}$ for a velocity $v$ of the inclusion;
while the latter is of the order of the top cylindrical area $\pi a_\mathrm{p}^2$ times the shear stress in the fluid, $v \eta/a_\mathrm{p}$.
When $\epsilon \gg 1$, the hydrodynamic resistance due to the motion 
in the membrane dominates; this is the regime considered by Saffman and Delbr\"{u}ck, which is relevant for small inclusions 
($\it i.e.$ for small values of the radius of the inclusion $a_p$, which is the only lateral length
scale considered in the model). 
Instead, when $\epsilon \ll 1$, the resistance occurs mainly due to the motion in the external fluid; 
this is the regime relevant for large inclusions. 
In this case, since the motion mainly occurs in the bulk external fluid, it is
described by the Stokes-Einstein formula, with a mobility going as $1/a_\mathrm{p}$. We therefore expect a cross-over between both regimes, 
when the size of the inclusion is of the order of $500$~nm, which is the estimate for $\ell$ for a typical ratio of viscosities $\eta_\mathrm{m}/\eta\simeq 100$ and a membrane thickness $h\simeq 5$~nm.

\section{Experimental tests of the Saffman-Delbr\"{u}ck model} \label{sec:Exp-SD}

Theoretically, a more complete solution of the SD hydrodynamic problem has been proposed which interpolates between the SD regime and the Stokes-Einstein regime \cite{Hughes1981}. 
Such expressions have been tested 
using simulations \cite{Guigas2008}, and later improved by the group of P. Schwille \cite{Petrov2008} who also carried out a set of careful experiments 
with micron-sized solid domains in giant unilamellar vesicles, confirming the results expected in the regime $\epsilon \ll 1$ \cite{Petrov2012}.

\begin{figure}[b]
\begin{center}
\includegraphics[scale=0.4]{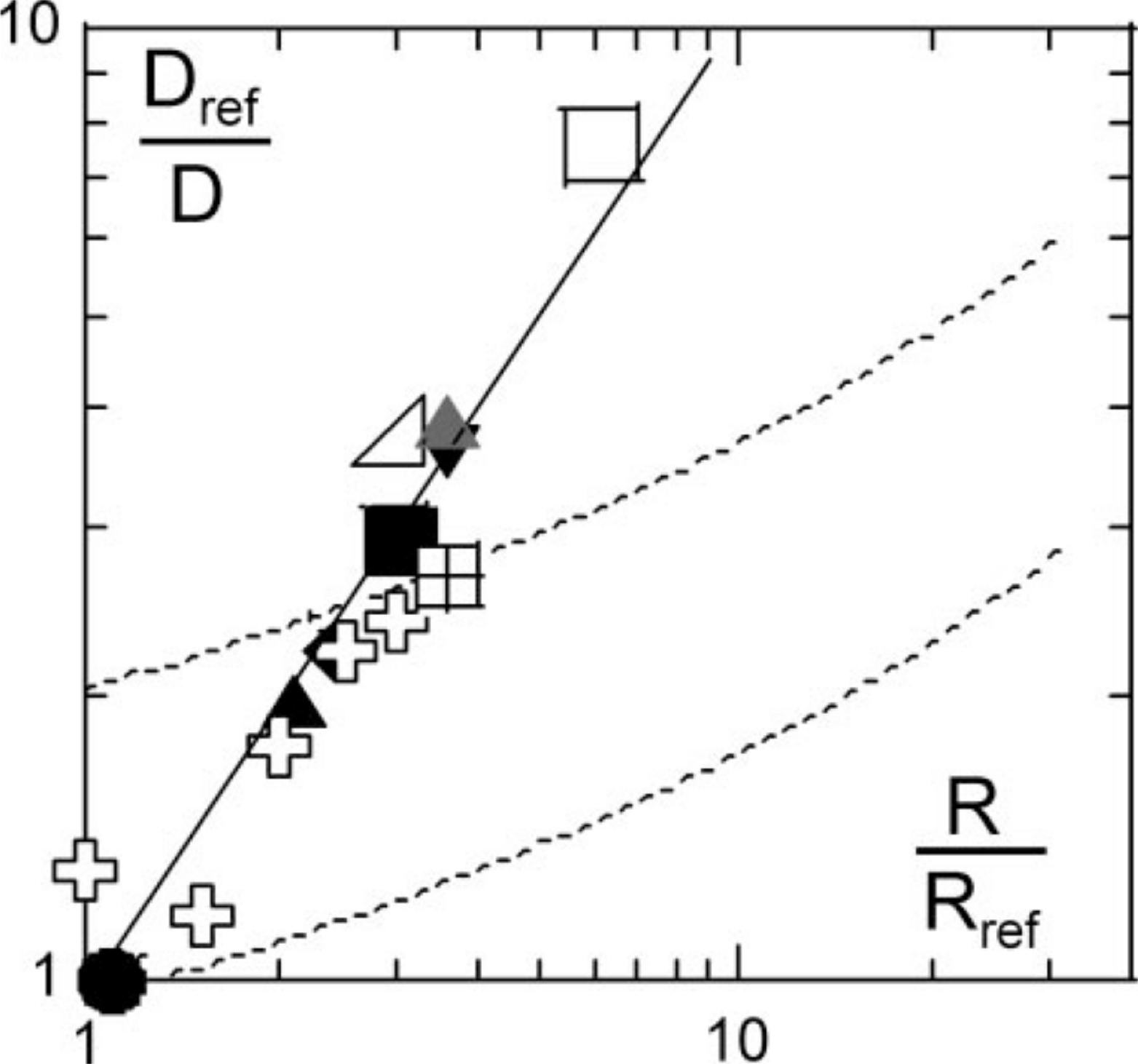}
\end{center}
%
%
\caption{Figure taken from Ref. \cite{Gambin2006}: Normalized inverse diffusion coefficient $D_{ref}/D$ 
vs. object radius $R/R_{ref}$, with open symbols representing data gathered from the literature, and filled symbols
data from Ref. \cite{Gambin2006}. Filled symbols represent peptides assemblies, 
with the peptide called $L_{12}$ as reference,  
while for oligomers of peptides (crosses), acetylcholine receptor
(AChR), bacteriorhodopsin (BR), and SR-ATPase (squares), the lipid diffusion serves as reference.
The solid line is a power-law regression leading to $D_{ref}/D \sim R^{1.04}$. For comparison,
the dashed line represents the prediction of the Saffman–Delbr\"{u}ck
model (Eq. 1) (upper line, same as in Fig. 3, and lower fit as in Fig. 2 of Ref. \cite{Gambin2006}).}
\label{fig:1}       
\end{figure}

Below, we focus on some of the experiments which have tested or challenged the SD model in 
the regime of small inclusions $\epsilon \gg 1$ and for flat membranes, which are conditions for which the model 
 should be applicable.
In 2006, Gambin et al.~\cite{Gambin2006} performed experiments with a large panel of peptides and membrane proteins with various shapes and particle sizes ranging from 5~nm to 30~nm; the diffusion coefficient has been measured using fringe pattern photobleaching. 
They reported a dependence of the diffusion coefficient on the size of the particle of the Stokes type ($D_0\sim1/a\ind{p}$, see Fig.~\ref{fig:1}), much stronger than the logarithmic dependence predicted by Eq. \ref{SD-D}.   
In a second set of experiments, 
Gambin et al. tuned the membrane thickness by swelling the membrane 
with an hydrophobic solvent, and measured the effect on the mobility of the peptides. 
They found that the mobility of the inclusions is maximal when their height matches
the membrane thickness. 
They attributed the reduced mobility of peptides with a smaller length than the bilayer thickness 
to the pinching of the bilayer (provided that the peptides are sufficiently long to span the bilayer).
On the other hand, peptides much longer than the membrane thickness cannot fit in the upright position and must tilt with respect to the membrane normal, which may cause, again, a reduction of their mobility.
These experiments triggered a lot of activity on the theoretical side, which we review in the theory section.

\begin{figure}[h]
\begin{center}
\includegraphics[width=0.7\linewidth]{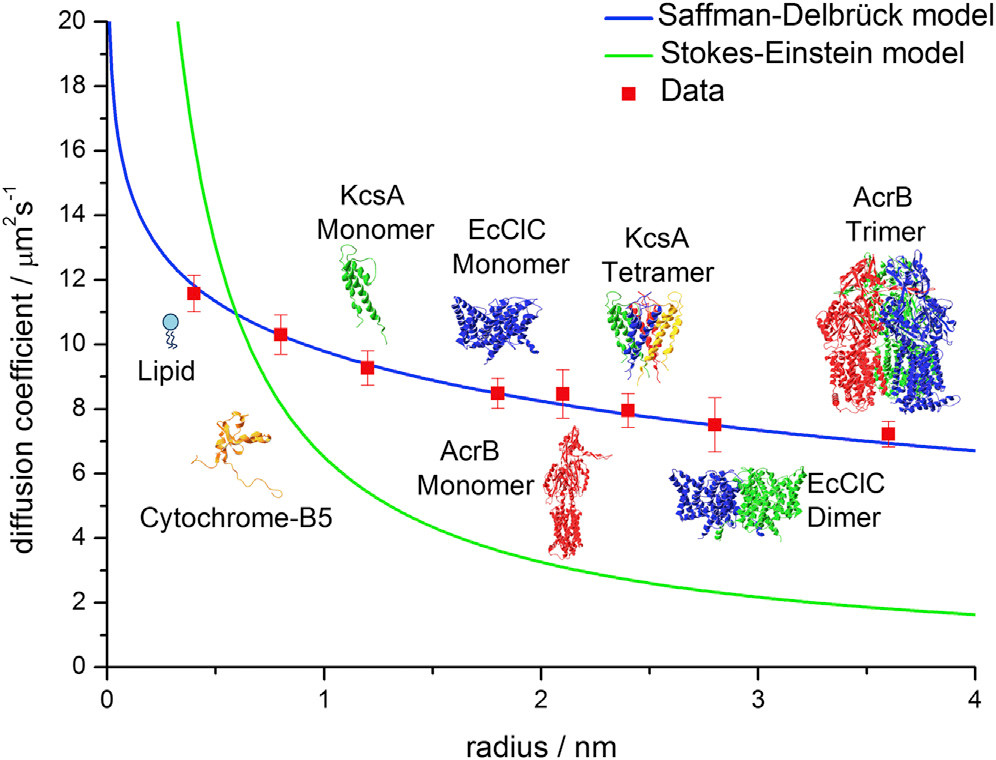}
\end{center}
\caption{Figure taken from Ref. \cite{Weis2013}: Saffman–Delbr\"{u}ck versus Stokes-Einstein model. The investigated
species are DPPE, cytochrome B5 (depicted without transmembrane
domain), KcsA, EcClC, and AcrB. The monomeric forms of membrane
proteins were directly added to the BLM. The oligomeric forms were reconstituted
via SNARE-mediated vesicle fusion. DPPE was labeled with
Atto655, and all proteins were labeled with Alexa647. In addition, we fitted
the HPW-based model suggested by Petrov and Schwille \cite{Petrov2008}, which can
reproduce the classical Saffman–Delbr\"{u}ck model in the size range investigated.
The fit parameter for all fits was the product of membrane viscosity
and thickness, $\eta_\mathrm{m} h$. The temperature was set to 295 K and the viscosity of
the surrounding buffer was $\eta= 0.96$ mPa.
}
\label{fig:weis}       
\end{figure}

In contrast to these experiments, one work \cite{Ramadurai2009} reported a weak dependence of $D_{0}$ on the protein radius 
$a\ind{p}$, using fluorescence correlation spectroscopy (FCS) in line with the predictions of the SD model from Eq. \ref{SD-D}. 
The discrepancy with the data of Gambin et al. is likely to come from differences in the experimental techniques which have been used in both cases.
Recently, Weiss et al. performed experiments with dual focus fluorescence correlation spectroscopy \cite{Weis2013}, 
which provided accurate measurements, of higher quality than with simple FCS. 
These more accurate measurements of diffusion coefficients for membrane proteins in black lipid membranes are in perfect agreement with the SD model. 
The original figure of Ref.~\cite{Weis2013} is reproduced here in Fig. \ref{fig:weis} with courtesy of J. Enderlein.  
We believe that the reason for this perfect agreement may be that these experiments have been performed with black lipid membranes, which are tense membranes.

The idea that the membrane tension may affect the protein mobility, has only been investigated recently in a study involving 
one of us \cite{Quemeneur2014}.  
In this work, the mobility of two transmembrane proteins with the same lateral size, aquaporin 0 (AQP0) and a
voltage-gated potassium channel (KvAP), have been measured by attaching quantum dots to these proteins and 
by tracking them. One advantage of using this technique of single particle tracking is that it is 
free of the possible artifacts due to averaging over a population of interacting proteins 
(on which the experiments reported in Refs.~\cite{Gambin2006,Gambin2010} rely). 
Whereas AQP0 does not deform the bilayer, KvAP is curved and bends the membrane. 
These experiments have shown that the curvature-coupled protein KvAP undergoes a significant increase
in mobility under tension -- an effect clearly beyond the SD model -- , whereas the mobility of the
curvature-neutral water channel AQP0 is insensitive to it and follows the prediction of the SD model, as shown in Fig. \ref{fig:2}. 
Importantly, at high tension, the mobility of both proteins agree well with the prediction of the SD model, which gives 
$D_0=2.5$~$\mu\mathrm{m}^2/\mathrm{s}$~\cite{Gambin2006}.

So far, we have only discussed proteins diffusing in flat membranes. However, the SD model has also been tested using the same technique 
of single particle tracking and the same protein KvAP, but in different membrane environments namely in the curved and confined space of membrane nanotubes 
 \cite{Domanov2011}. In this work, it was found that measurements of the mobility of this protein in the tube are 
well described by an extension of the SD model to cylindrical geometries \cite{Daniels2007}.

Finally, let us mention some applications of the SD model for microrheology. Similarly to the Einstein relation, which allows the determination of 
the 3D local viscosity, based on measurements of the fluctuations of one or two tracers in a bulk fluid, the SD relation allows the determination of 
local membrane viscosity from measurements of the fluctuations of a tracer. This tracer can be a protein, but in that case it is crucial 
to properly understand the coupling between the protein 
and the membrane, as emphasized in the present chapter. In order to disentangle effects due to the tracer and the membrane, it is 
helpful if possible to combine measurements of translation or rotation diffusion as shown in Ref. \cite{Hormel2014}.

\section{Relevant geometrical or mechanical factors and protein-membrane couplings} \label{sec:Coupling}

Various geometrical or mechanical factors can affect the mobility of membrane proteins. The main factors are listed below:
\begin{itemize}
\item Geometrical properties of the membrane. The most basic geometrical parameter of the membrane is its thickness. For 
lipid membranes, this parameter is generally assumed to be constant of the order of 4 to 5nm. 
However, biological membranes are intrinsically heterogeneous and usually composed of mixtures of several kinds of lipid domains. 
In such systems, the membrane thickness can vary depending on the location inside or outside the 
domains.  Another very important local geometrical property of the membrane is its curvature. A preference of certain membrane proteins 
for membrane curvature means in particular that it is possible 
to sort them according to the local curvature as demonstrated in Ref. \cite{Aimon2014}.

\item Mechanical and chemical properties of the membrane. The former ones include the elastic moduli of the membrane, 
such as its bending modulus and its tension, and dissipative moduli, such as the membrane
viscosity and the relative friction between membrane leaflets. 
The latter ones include the chemical composition of the membrane, which controls many of its properties.
Both parameters should be regarded as local or global depending on the level of heterogeneity.

\item Geometrical parameters and mechanical properties of the membrane protein. 
The geometrical parameters of the protein include its size and shape~\cite{Chang1998,Liu2009,Camley2013,Aimon2014}.
Its shape notably defines its spontaneous curvature, which is a scalar for isotropic inclusions but becomes a tensor for anisotropic ones.
Other important protein mechanical properties include its compliance, 
which can be split into elastic and dissipative moduli. 
\item The solvent in which the protein and the membrane are embedded. It is mainly characterized by its 
viscosity which controls the hydrodynamic part of the dissipation. 
In the case that the membrane and its solvent are confined by rigid walls, the solvent will be also
described by its thickness, which can affect the mobility of membrane proteins 
via hydrodynamic screening effects \cite{Seki2011,Stone1998}.
\item Other inclusions present in the membrane modify the mobility of a given membrane protein, 
via direct interactions such as exclusion or 
indirect ones, such as membrane mediated interactions~\cite{Goulian1993,Reynwar2008}. Since at the microscopic scale, biological membranes are a crowded mix of membrane proteins and lipid partners, such collective 
effects are expected to be important.
\end{itemize}

In order to understand the way mechanical factors affect the mobility of a given protein, it is important to
focus on the mechanism by which the protein couples to the membrane.
The main coupling mechanisms are:
\begin{itemize}
\item Hydrodynamic coupling. Since the protein evolves in the membrane which is fluid, there is a clear hydrodynamic coupling due to the fluid membrane. But since the external fluid may be dragged by the motion of the lipids induced by the protein, there is also hydrodynamic coupling with the external fluid, which is usually water. 
This is the coupling considered by Saffman and Delbrück~\cite{Saffman1975} and in \cite{Seki2011,Stone1998}. 

\item Coupling to the local membrane curvature. A mismatch between the spontaneous curvature of the protein and the curvature of the membrane deforms the lipids around the protein, leading to an energetic penalty~\cite{Goulian1993,Reister2005,Reynwar2007,Naji2009,Quemeneur2014}. 
The precise form of this coupling depends whether the protein is assumed to be hard or soft. 

\item Coupling to the membrane thickness. A hydrophobic mismatch between the protein and the membrane thickness stretches 
or compresses the lipid tails around the protein~\cite{Andersen2007}.

\item Coupling to the membrane composition. Such an effect will be present when the protein 
has a special affinity for one kind of lipids while the membrane is made of a mix of various lipids~\cite{Reynwar2008}.

\item Electrostatic coupling. Various mechanisms are possible, depending whether the protein and the lipids are charged or not. 
Note that even if the lipids are uncharged, such couplings may be present since the lipids may be polarized locally by the protein if the later is charged. For membrane ion channels, 
there is generally a coupling of electrostatic origin in the presence of a transmembrane voltage since the field
lines are deformed locally near the protein \cite{Winterhalter1987}. Such a coupling is involved in the mechanism of voltage gating 
in ion channels \cite{Phillips2009}.  

\item Coupling due to geometrical effects. This form of coupling arises because the real trajectories of 
membrane proteins occur in 3D whereas these trajectories are typically recorded experimentally in a 
2D space~\cite{Reister2005,Reister-Gottfried2007}. 
This projection of real trajectories on 2D space results in an additional reduction of the protein mobility.
\end{itemize}

Naturally, a given protein may couple to the membrane via several couplings of this kind simultaneously, and more 
couplings are possible. 
In the next section, we explore non-hydrodynamic couplings, and their consequences for the mobility of the membrane protein.  

\section{Theoretical models} \label{sec:Theory}

SD theory implicitly assumes that the protein diffuses in a membrane which remains flat and unaffected by the presence of the protein. 
Therefore, a possible origin for the discrepancy observed by Gambin et al. \cite{Gambin2006} is the significant 
local membrane deformation due to the interaction between the protein and the lipid bilayer as 
 proposed in 2007 by Naji et al. \cite{Naji2007}. In this view, a given membrane protein 
should experience additional dissipation, 
either within the membrane or within the external fluid, due to the local deformation 
which it carries along as it diffuses in the membrane.

In the next subsection, we sketch the original theoretical argument put forward by Naji et al.
Then, we turn to a more formal analysis of the drag coefficient for a general  
order parameter using the so called polaron model. In the next subsection, this polaron model is applied to 
the specific membrane curvature coupling and used to analyze the experiments of Ref. \cite{Quemeneur2014}.
We finish with other potential applications of this framework and with a list of open problems.

\subsection{Heuristic approach to the membrane perturbation}\label{sub:heuristic_polaron}

Naji et al. suggested that the discrepancy between the SD prediction~\cite{Saffman1975} and the experimental results of Gambin et al.~\cite{Gambin2006} could be attributed to the perturbation of a local order parameter $\phi(\rv)$ of the membrane that could represent its lipid composition, thickness, or height~\cite{Naji2007}.
They assumed that the perturbation of the order parameter $\phi$ has a characteristic length $\xi$, and is dragged along with the protein, 
dissipating energy in a boundary layer of width $\delta \xi$. 
The power dissipated by this process is $P_\phi\sim 2\pi(a\ind{p}+\xi)\delta\xi v^2$, corresponding to a drag coefficient
\begin{equation}
\lambda_\phi\sim 2\pi(a\ind{p}+\xi)\delta\xi.
\end{equation}
If the length scale of the perturbation is small with respect to the size of the protein, $\xi \ll a\ind{p}$, the drag coefficient takes the form $\lambda_\phi =\Gamma a\ind{p}$.
This drag coefficient enters the Einstein relation for the diffusion coefficient together with the SD drag $\lambda\ind{SD}$, giving
\begin{equation}
D=\frac{kT}{\lambda\ind{SD}+\lambda_\phi}=\frac{kT}{\lambda\ind{SD}+\Gamma a\ind{p}}.
\end{equation}
If the dissipation due to the perturbation of the order parameter $\phi$ dominates, the diffusion coefficient scales as $D\sim 1/a\ind{p}$, which is compatible with the experiments of Gambin et al. \cite{Gambin2006}.

In the following subsection, we turn to a formal description of the perturbation of the order parameter 
$\phi$, which allows to compute the drag coefficient precisely.

\subsection{A polaron model for the perturbation of an order parameter}\label{sec:polaron}

The back-action on an object that modifies its environment generates a drag force known as the polaron effect, 
which was originally described for an electron moving in a lattice~\cite{Landau1933}. 
A polaron is a charge carrier which deforms a surrounding lattice and moves in it with an induced polarization field.
This effect has been applied to soft matter 
systems by one of us~\cite{Demery2010,Demery2010a}.
We recall briefly the steps that allow to compute the drag coefficient generated by the coupling between the protein and a local 
order parameter (or field) $\phi(\rv)$, which may represent, e.g., the membrane height, thickness, or composition.


%

\subsubsection{Definition of the model}\label{}

There are two steps in the modeling of the coupling between the protein and the order parameter $\phi(\rv)$ 
of the membrane. The first is to write the Hamiltonian of the protein-order parameter system, that we assume to be of the form
\begin{equation}\label{eq:polaron_energy}
\Ha[\phi(\rv),\R]=\frac{1}{2}\int \phi(\rv)\Delta*\phi(\rv)d^2\rv -  K*\phi(\R),
\end{equation}
where the star denotes the convolution product $A*B(\rv)=\int A(\rv-\rv')B(\rv')d^2\rv'$.

The first term is the energy of the order parameter itself, it is quadratic in the field and determined completely by the operator $\Delta(\rv)$. The second term couples the order parameter to the position of the protein, it is linear in the field and determined by the operator $K(\rv)$.
The second step is to give the dynamics of the field, that we assume to be overdamped,
\begin{equation} \label{eq:polaron_dynamics}
\frac{\partial \phi}{\partial t}(\rv,t) = - R*\frac{\delta\Ha}{\delta\phi(\rv,t)} + \eta(\rv,t),
\end{equation}
where $R(\rv)$ is a linear operator and $\eta(\rv,t)$ is a Gaussian white noise with correlation function
\begin{equation}
\left\langle \eta(\rv,t)\eta(\rv',t') \right\rangle = 2T R(\rv-\rv')\delta(t-t'),
\end{equation}
and $\delta\Ha / \delta\phi(\cdot,t)$ denotes a functional derivative \cite{Chaikin1995_vol}.

The presence of the operator $R(\rv)$ in the noise correlation ensures that detailed balance is satisfied.
Under this form, the model is completely determined by the three linear operators $\Delta(\rv)$, $K(\rv)$ and $R(\rv)$.

\subsubsection{Computation of the drag coefficient}\label{sec:computation_drag_coeff}

The force exerted by the field on the protein is given by the gradient of the interaction energy,
\begin{equation}\label{eq:def_force_polaron}
\fv[\phi(\rv),\R] = -\nabla_\R\Ha[\phi(\rv),\R] = \nabla K*\phi(\R).
\end{equation}

To compute the drag coefficient, a constant velocity $\vv$ is imposed to the protein, $\R(t)=\vv t$. 
We introduce the average field in the reference frame of the protein, which is time-independent in the stationary regime,
\begin{equation}
\Phi(\rv) = \langle \phi(\rv-\R(t),t) \rangle,
\end{equation}
The average drag force can be written as
\begin{equation}
\left\langle \fv[\phi(\rv,t),\R(t)] \right\rangle = \nabla K*\Phi(0).
\end{equation}

The average field in the reference frame of the protein is obtained by averaging Eq.~\ref{eq:polaron_dynamics} in the moving frame:
\begin{equation}
-\vv\cdot\nabla\Phi(\rv) = -R*\Delta*\Phi(\rv) + R*K(-\rv),
\end{equation}
This equation is solved in Fourier space,
\begin{equation}
\tilde\Phi(\kv)=\int e^{-i\kv\cdot\rv}\Phi(\rv)d^2\rv = \frac{\tilde R(\kv)\tilde K^*(\kv)}{\tilde R(\kv)\tilde\Delta(\kv)-i\kv\cdot\vv}.
\end{equation}
The force can be deduced from this Fourier transform,
\begin{equation}
\langle \fv \rangle = \int i\kv \tilde K(\kv)\tilde\Phi(\kv)\frac{d^2\kv}{(2\pi)^2} = \int \frac{ i\kv\tilde R(\kv)\left|\tilde K(\kv)\right|^2}{\tilde R(\kv)\tilde\Delta(\kv)-i\kv\cdot\vv}\frac{d^2\kv}{(2\pi)^2}  
\end{equation}
Expanding the force at small velocity leads to
\begin{equation}
\langle \fv \rangle \underset{|\vv|\to 0}{\sim} -\frac{\vv}{2} \int \frac{ \kv^2\left|\tilde K(\kv)\right|^2}{\tilde R(\kv)\tilde\Delta(\kv)^2}\frac{d^2\kv}{(2\pi)^2},
\end{equation}
where we used that $\int \kv\cdot\vv \kv f(|\kv|)d^2\kv = \frac{1}{2}\vv\int\kv^2f(|\kv|)d^2\kv$ for any function $f(|\kv|)$.
The drag coefficient is thus
\begin{equation}\label{eq:polaron_drag}
\lambda = \frac{1}{2} \int \frac{ \kv^2\left|\tilde K(\kv)\right|^2}{\tilde R(\kv)\tilde\Delta(\kv)^2}\frac{d^2\kv}{(2\pi)^2}
=\frac{1}{4\pi}\int_0^\infty \frac{k^3 \left|\tilde K(k) \right|^2}{\tilde R(k)\tilde\Delta(k)^2}dk.
\end{equation}
The second equality assumes isotropic operators.

If this integral diverges at large wavevectors $k$, the finite size $a\ind{p}>0$ of the protein can be used to cut-off the integral at $k\ind{max}=\pi/a\ind{p}$~\cite{Demery2010,Demery2010a}. This regularization introduces a dependence of the drag on the size of the protein, whose form depends on the divergence of the integral.



The formula (\ref{eq:polaron_drag}) involves the three operators that determine the model.
We shall see below how it applies to the coupling of the membrane curvature with a protein with spontaneous curvature.

\subsubsection{Link to the diffusion coefficient}\label{sec:link_drag_diffusion}

The Einstein relation~\cite{Einstein1905} relates the diffusion coefficient $D$ to the drag coefficient $\lambda_\mathrm{f}$ defined as the ratio between the constant force applied to the particle and its average velocity, $\fv=\lambda_\mathrm{f} \langle \vv \rangle$ by $D=k\ind{B}T/\lambda\ind{f}$.
In a fluctuating field, the drag coefficient $\lambda\ind{f}$ and the drag coefficient $\lambda$ computed \emph{at constant velocity} are not equal~\cite{Dean2011,Demery2011}.
However, in the adiabatic regime where the field equilibrates much faster than the particle, the field remains close to equilibrium and the two drag coefficients are equal~\cite{Dean2011,Demery2011}. 
In this case, the diffusion coefficient can be written as
\begin{equation}\label{eq:einstein_adia}
D\ind{eff}=\frac{k\ind{B}T}{\lambda\ind{SD}+\lambda_\phi},
\end{equation}
where $\lambda_\phi$ is given by Eq.~\ref{eq:polaron_drag}.

\subsection{Application to a protein that couples to the membrane curvature}

The coupling of a protein to the membrane curvature and its effect on its diffusion coefficient has been investigated before the introduction of the general model presented above~\cite{Goulian1993,Naji2009, Reister2005,Leitenberger2008,Reister_Gottfried2010}.
These models resemble the model above where the operators have been specified for the coupling to the membrane curvature; the differences are discussed in Sec.~\ref{sec:extension_polaron}.


Let us consider a single protein diffusing on a membrane patch of size $L\gg a_\mathrm{p}$ described by a height function $h(\textbf{r})$. We use 
the modified Helfrich Hamiltonian:
\begin{equation}
\Ha_0[h,\R]=\frac{\kappa}{2} \int d^2 \rv \left[ \left(\nabla^2 h \right)^2 +
\frac{\Sigma}{\kappa} \left( \nabla h \right) ^2 - \Theta G(\mathbf{r}-\R) \nabla^2 h \right],
\label{HS_lin}
\end{equation}
where the first two terms represent the energy of elastic bending of the bilayer with modulus $\kappa$ and tension $\Sigma$, and the 
last term models the membrane curvature induced at the location of the protein $\R$, which is time-dependent. 
The strength of the induced curvature scales linearly with the protein spontaneous curvature $C_p$, $\Theta = 4 \pi a_{p}^{2}C_p$, 
similarly to~\cite{Naji2007,Naji2009}. The range of influence of the protein on the membrane is modeled by the weight function 
$G$ which is normalized to one and is non-zero over a distance of the order of $a_\mathrm{p}$. This Hamiltonian carries with it a cutoff length $a$, which corresponds to the bilayer thickness ($\sim$ 5 nm).  
This model is a particular case of the polaron model described in the previous section, where the operators are given in Fourier space by
\begin{align}
\tilde{\Delta}(k) & = \kappa k^4 + \Sigma k^2, \\
\tilde{R}(k)      & =  \frac{1}{4 \eta k}, \\
\tilde{K}(\kv)    & =- \frac{\kappa \Theta}{2} k^2 \tilde{G}(\kv).
\end{align}

As explained in Sec.~\ref{sec:link_drag_diffusion}, the diffusion coefficient is given by Eq.~\ref{eq:einstein_adia} if the dynamics of the membrane is much faster than the diffusion of the inclusion.
This adiabatic approximation has been checked using numerical simulations \cite{Naji2009} and by an explicit evaluation of the slowest membrane relaxation times for the conditions of the experiment of Ref.~\cite{Quemeneur2014}. 

\begin{figure}[h]
\begin{center}
\includegraphics[scale=1]{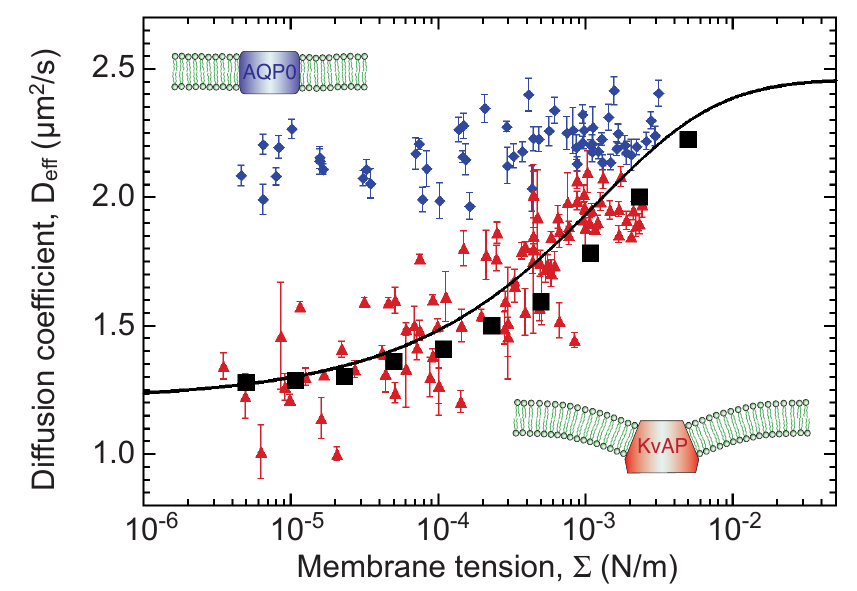}
\end{center}
%
%
\caption{Figure taken from Ref. \cite{Quemeneur2014}: Protein lateral mobility in fluctuating membranes. 
Semi-logarithmic plot of the diffusion coefficients ($D_\mathrm{eff}$) as a function of the membrane tension 
$\Sigma$, for AQP0 (\textcolor{blue}{$\blacklozenge$}) 
and KvAP (\textcolor{red}{$\blacktriangle$}) 
labeled with streptavidin QDs. 
KvAP data adjusted by Eq. \ref{effective D_lin} (solid line) yields a protein coupling coefficient $\Theta$ = 3.5$\times10^{-7}$ m considering  $a = 5$ nm, 
$\kappa = 20$ k$_B$T and $D_0 \simeq$ 2.5 
m$^2$/s. Simulations of the protein diffusion on a membrane subject to thermal fluctuations 
($\blacksquare$) 
agree well with the experimental data and theory. Insets: sketches of membrane deformation near proteins.}
\label{fig:2}       
\end{figure}

The polaron approach thus provides explicit predictions for the membrane profile and for the effective friction coefficient of the protein. 
The effective friction coefficient is 
 $\lambda(\sigma)=\Theta^2 \eta W_0(\sigma)/2 a$ where $\sigma = 
\Sigma a^2/ 4 \pi \kappa$ is a reduced tension and $W_0(\sigma)$ a function given in Eq.~S27 of Ref. \cite{Quemeneur2014}. 
The diffusion coefficient $D_\mathrm{eff}$ is obtained from the effective drag coefficient using the Stokes-Einstein relation:
\be
\frac{D_0}{D_\mathrm{eff}}=1 + \frac{\eta D_0 \Theta^2 W_0(\sigma)}{2a k_\text{B} T},
\label{effective D_lin}
\ee
where $D_0$ represents the bare diffusion coefficient of the inclusion in a flat tense membrane, given by Eq.~\ref{SD-D}. The tension dependence 
of $D_\mathrm{eff}$ is shown in Fig. \ref{fig:2}, together with the experimental data and the 
simulation data. For the simulation data, the protein diffusivity 
was directly computed from the MSD of stochastic trajectories, which were generated by numerically integrating 
the stochastic equations of motion of 
the inclusion and the membrane. 

\begin{figure}[h]
\begin{center}
\includegraphics[width=8.5cm]{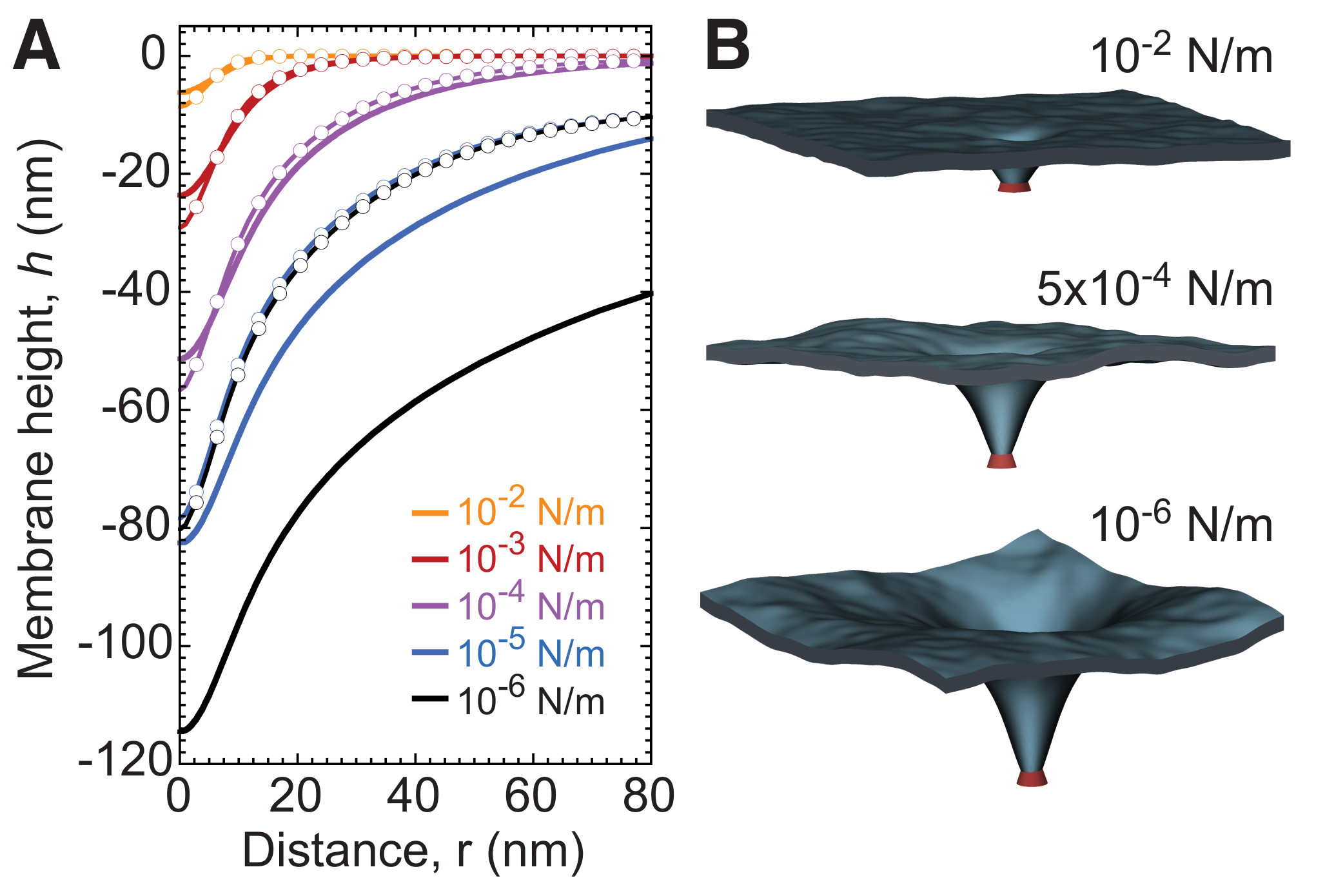}
\end{center}
%
%
\caption{Figure taken from Ref. \cite{Quemeneur2014}: Membrane shape as a function of the tension. (A) Theoretical profile with parameters $\Theta = 3.5\times10^{-7}$m, $\kappa$ = 20 k$_B$T, $a_\mathrm{p}$ = 4 nm (solid lines). Profile calculated from the numerical energy minimum shape using simulation (dashed line).
 The height value far away from the inclusion is chosen to be zero. Discrepancy at low tension originates from the finite size of the membrane 
$L^2$ used in the simulations (see SI). (B) Three-dimensional membrane profiles obtained from numerical simulations 
for three different tensions.}
\label{fig:3}       
\end{figure}

The membrane profile around the inclusion is obtained by the same method. 
The lateral characteristic width of this profile is the cross-over length between the tension 
and the bending regime for the fluctuations, namely $\xi=\sqrt{\kappa/\Sigma}$, while the characteristic height of the membrane 
deformation at zero tension scales as $\Theta$. The geometry of the local deformation from the membrane mid-plane 
induced by KvAP when subjected to various tensions is shown in Fig. \ref{fig:3}. Using the method of 
\cite{Naji2007,Naji2009,Sigurdsson2013}, we have carried out simulations, which also confirm
 the expected theoretical membrane profile as shown in Fig. \ref{fig:3}.

\subsubsection{Discussion of the results of the polaron model}

We find that theory and simulations fit very well the experimental results for the diffusion coefficient of KvAP using a coupling coefficient $\Theta = 3.5\times10^{-7}$ m. This proves that the mechanical coupling between the proteins and the membrane can strongly affect protein mobility. At high tension, 
the experimental data of AQP0 and KvAP converge to the same plateau value, consistently with the SD limit since they have about the 
same steric radius. As a control, the diffusion coefficient of pure lipids was measured and found to be independent of tension as for AQP0. 
The corresponding constant value for $D_0$ agrees with the prediction of the SD model, using a lipid size $a_\mathrm{p}=0.5$ nm. 
At lower tension, Fig. \ref{fig:2} shows that the KvAP data starts to deviate significantly from the plateau at $\Sigma \sim 5 \times 10^{-3}$ N/m, 
which corresponds to the point at which the lateral characteristic length $\xi$ is of the order of the protein size.


An additional outcome of this approach concerns the dependence of $D_\mathrm{eff}$ with the protein radius $a_\mathrm{p}$, which can be obtained from the following scaling argument. Since the protein makes a fixed angle with respect to the membrane, $C_p \sim 1/a_\mathrm{p}$. Given the relation $\Theta = 4 \pi a_{p}^{2}C_p$, this implies $\Theta \sim a_\mathrm{p}$. Below the cross-over to the SD regime, the local membrane deformation is much larger than the protein size $a_\mathrm{p} \ll \xi$, and the drag is dominated by the contribution due to the membrane deformation. Therefore using Eq.~\ref{effective D_lin}, one finds $D_\mathrm{eff} \sim k_B T a/a_\mathrm{p}^2$, in agreement with Ref.~\cite{Demery2010}. Note that such a result is also compatible with the Stokes-Einstein scaling law in $1/a_\mathrm{p}$ obtained in Ref.~\cite{Naji2007}, because in this reference only one characteristic length for the protein is used thus $a \simeq a_\mathrm{p}$. 

Despite a good agreement between model and data, the physical interpretation of the coupling coefficient $\Theta$ requires a more detailed discussion. Indeed, the spontaneous curvature deduced from this coupling coefficient via a fit of the data is significantly larger than that obtained from thermodynamic measurements, based on the preferential sorting at equilibrium of the proteins between GUV and highly curved membrane nanotubes \cite{Aimon2014}. One possible interpretation for this discrepancy is that in dynamic measurements, the basic relevant object, namely the association of the moving protein with the deformed membrane around it, may have a size larger than $a_\mathrm{p}$. Such an enhancement of the size could in principle describe physically a layer of lipids dragged by the motion of the protein as considered in \cite{Camley2012}. However, given the value of the coupling coefficient $\Theta$, 
this translates into an effective radius of 47 nm, a rather large value with respect to the lipid and protein sizes (0.5 and 4 nm respectively). 
For this reason, we have proposed in Ref. \cite{Quemeneur2014} an alternate explanation, namely that this discrepancy reflects an additional source 
of internal dissipation, which for the present problem, could arise from inter-monolayer slip due to the motion of the inclusion. 
By considering this additional dissipative mechanism internal to the membrane, we have shown that we can still account for the dependence of $D_\mathrm{eff}$ versus $\Sigma$, but with a lower a coupling coefficient $\Theta = 3.4\times10^{-8}$ m, corresponding to a $C_p$ = 0.16 nm$^{-1}$. This value is then much more compatible with the thermodynamic measurements previously reported in \cite{Aimon2014}.

\subsection{Application to other order parameters}\label{sec:other_order_parameters}

An important advantage of the polaron model is that it can be applied to other order parameters than the membrane height.
Already in the previous subsection, we have mentioned an extension of the original polaron model, including dissipation mechanisms 
 internal to the membrane, as a possible scenario to explain the value of the coupling constant in Ref. \cite{Quemeneur2014}. 
In that extension, the relevant order parameter was already no longer the membrane height field, but rather the difference of lipid 
densities in the two leaflets of the membrane.
Below, we explain how different order parameters can be described in the polaron model.

\subsubsection{Coupling to the thickness}\label{}

When the hydrophobic length of the protein differs from the membrane height, the membrane is compressed or stretched close to the protein. The order parameter 
$\phi(\rv)$ is the difference between the local thickness of the membrane and its average thickness.
If the membrane has a bending modulus $\kappa$ and a compressibility $\chi$, the energetic operators are given by~\cite{Gruner1985,Mouritsen1993,Andersen2007}:
\begin{align}
\tilde\Delta(k) & = \kappa k^4 + \chi,\\
\tilde K(k) & = \pi a\ind{p} \Theta,
\end{align}
where the coupling constant $\Theta$ now depends on the height mismatch and the energetic cost for exposing an hydrophobic area.

To our knowledge, no dynamics has been proposed for the system. However, if we assume a simple form  
for the membrane friction (defined by the parameter $\gamma$ in a way which conforms to model A dynamics \cite{Chaikin1995_vol})  
and if we add to that the usual hydrodynamic friction due to the flow created in the solvent, we can propose an operator of the form
\begin{equation}
\tilde R(k) = \gamma + \frac{1}{4\eta k}.
\end{equation}

\subsubsection{Coupling to the composition}\label{}

Real membranes are composed of several types of lipids, which may be mixed or separated. Proteins may couple to the composition field if they are preferentially wetted by one kind of lipids. 
Close to the demixing transition, which can be controlled \emph{in vitro} in lipid vesicles~\cite{Veatch2003,Veatch2007},
 this coupling to the composition induces long-range forces between the proteins~\cite{Machta2012}. Numerical simulations have shown
 that it may induce the aggregation of proteins~\cite{Reynwar2008}.

In the simplest case, the relevant order parameter $\phi(\rv)$ is the difference between the local and average concentrations of a lipid specie. Its energy involves its correlation length, $\xi$, which diverges at the demixing transition, allowing long-range forces. The dynamics is conserved, the timescale being set by the diffusion coefficient $D_0$ of the lipids. The operators take the form~\cite{Taniguchi1996}
\begin{align}
\tilde\Delta(k) & = A\left[k^2 + \xi^{-2}\right],\\
\tilde K(k) & = \Theta,\\
\tilde R(k) & = D_0k^2,
\end{align}
where $\Theta$ quantifies again the strength of the coupling.
Similarly to what happens for the interaction between proteins, the strongest effect on the mobility 
is expected close to the demixing transition (see also Ref.~\cite{Fujitani2013} for a careful calculation of 
the drag coefficient of a diffusing domain in a near critical binary mixture).

In the description given above, we have not included the fact that the different lipid species may have different spontaneous curvatures. 
If this is the case~\cite{Taniguchi1996,Yanagisawa2007}, the concentration field is coupled to the curvature field and there are
 two order parameters.

\subsubsection{Coupling to other fields}\label{}

The membrane could be coupled to other order parameters, and to several parameters at the same time. 
This situation has been evoked in the previous section, where the curvature is coupled to the composition.
Also, a coupling of the membrane curvature to the lipid density in the two leaflets has been suggested to provide another dissipative mechanism in Ref.~\cite{Quemeneur2014}. 
The polaron model needs to be extended to deal with these situations where there are several order parameters.

Another possible order parameter is the liquid crystalline order parameter of the lipids, which is expected to 
be distorted near small inclusions like proteins \cite{Bartolo2000}.
The dynamics of a liquid crystalline order parameter can be described using continuum theories of liquid crystals 
and there are many studies on the dynamics of topological defects in liquid crystals \cite{Chaikin1995_vol,Gennes1993}.  
Pre-transition (or pre-melting) effects which are well known in liquid crystals are also relevant 
for transmembrane proteins, which can for instance stabilize a microscopic order–disorder
interface, when their hydrophobic thickness matches that of the
disordered phase and when they are embedded in an ordered bilayer \cite{Katira2016}. 

It is important to appreciate that if the protein couples to the membrane through an order parameter different from the height field, like a concentration field or a nematic order parameter field, it could happen that there will be no observable membrane deformation near the protein, while the order parameter is still perturbed.


\subsection{Extension of the polaron model}\label{sec:extension_polaron}

We discuss here three ways to extend the simple polaron model presented in Sec.~\ref{sec:polaron}. 
First, we consider the case where several order parameters are coupled together and with the protein.
Second, we discuss different ways to couple the order parameter and the protein.
Finally, the coupling of the dynamics of the order parameter to the hydrodynamic flow in the membrane is evoked.

\subsubsection{Coupling to several order parameters}\label{}

There are cases where several order parameters $\phi_\alpha(\rv)$ are coupled~\cite{Taniguchi1996,Quemeneur2014}. 
The polaron model can be extended to handle several order parameters. The three operators become matrix operators:
\begin{align}
\Delta(\rv) & \rightarrow \Delta_{\alpha\beta}(\rv),\\
K(\rv) & \rightarrow K_\alpha(\rv),\\
R(\rv) & \rightarrow R_{\alpha\beta}(\rv).
\end{align}
In this way, the new hamiltonian becomes
\begin{equation}\label{eq:new hamiltonian}
\Ha[\phi(\rv),\R]=\frac{1}{2}\int \phi_\alpha(\rv) \Delta_{\alpha \beta} *\phi_\beta(\rv) d^2\rv -  K_\alpha*\phi_\alpha(\R),
\end{equation}
where the greek indices run over the different order parameters and a summation over repeated indices is implicit.

The extension of the computation of the drag coefficient given in Sec.~\ref{sec:computation_drag_coeff} is straightforward, leading to
\begin{equation}
\lambda = \frac{1}{2}\int \kv^2 \tilde K(\kv)_\alpha \tilde \Delta(\kv)^{-1}_{\alpha\beta} \tilde R(\kv)^{-1}_{\beta\gamma} \tilde \Delta(\kv)^{-1}_{\gamma\delta} \tilde K(\kv)_\delta^*\frac{d^2\kv}{(2\pi)^2}.
\end{equation}

\subsubsection{Other protein couplings to the order parameter}\label{}

In the example of a protein coupled to the curvature of the membrane, it is clear that a linear coupling cannot be used to ``impose'' a given curvature to the membrane at the location of the protein. Instead, the value of the membrane curvature depends on the Hamiltonian operators $\Delta(\rv)$ and $K(\rv)$.

The more direct way to impose the membrane curvature is to set it as a boundary condition~\cite{Bitbol2010,Camley2012}, using a relation of the form 
$K*\phi(\R)=A$ (in this case of a coupling to the membrane curvature, $A$ corresponds to the curvature of the membrane at the location of the protein, 
denoted $C_p$ above). In this framework the force can no longer be computed with Eq.~\ref{eq:def_force_polaron}, 
and one should instead integrate the stress tensor around the protein~\cite{Bitbol2011,Capovilla2002}.

An imposed boundary condition can be modeled in the framework of the polaron model, by introducing a parameter $\alpha$ 
in front of the coupling operator $K(\rv)$ and tuning it so that the boundary condition is satisfied. In this case, $\alpha$ depends on the other parameters of the model (e.g., the membrane tension and bending modulus when the protein is coupled to the curvature of the membrane).
This approach has been shown to give results consistent with those obtained with an imposed boundary condition~\cite{Camley2012}.

Alternatively, the polaron model can be modified by replacing the linear coupling to the order parameter by a coupling of the form
\begin{equation}\label{eq:polaron_quadratic}
\Ha\ind{int}[\phi(\rv),\R]=\frac{\kappa_\mathrm{p}}{2}\left[K*\phi(\R)-A \right]^2,
\end{equation}
where $\kappa_\mathrm{p}$ is the ``rigidity'' of the coupling (e.g., the bending rigidity of the protein).
In the limit of infinite rigidity, $\kappa_\mathrm{p} \to\infty$, one recovers the imposed boundary condition.
This form of coupling has been used to model the coupling to curvature in Refs.~\cite{Naji2009, Reister_Gottfried2010}.
In the limit where $K*\phi(\R)$ is small with respect to $A$, the linear coupling introduced in Eq.~\ref{eq:polaron_energy} is recovered.

In the particular case $A=0$, the coupling does not affect the average value of the order parameter, but only its fluctuations. This Casimir-like coupling gives rise to another source of drag~\cite{Demery2011a} and further reduction of the diffusion coefficient \cite{Demery2013}.
For a coupling to curvature, when both effects (average field deformation and fluctuations) compete, the fluctuations induced part is often neglected 
owing to the fact that the thermal fluctuations are small, i.e. $k\ind{B}T\ll\kappa$.

In the general case, the interaction Hamiltonian given in Eq.~\ref{eq:polaron_quadratic} depends on one more parameter than the linear interaction in Eq.~\ref{eq:polaron_energy}. 
This has important implications: for example, when the coupling to membrane curvature is modeled by a linear interaction, as in Eq.~\ref{HS_lin}, the rigidity of the protein and its spontaneous curvature cannot be decoupled. On the other hand the rigidity $\kappa_\mathrm{p}$ and 
the preferred average value $A$ have very different effects with the quadratic coupling, Eq.~\ref{eq:polaron_quadratic}.

\subsubsection{Interaction of the order parameter with the hydrodynamic flow}\label{label}

The polaron model is completely independent of the hydrodynamic calculation of Saffman and Delbrück, in the sense that the flow 
in the membrane and in the solvent are neglected, and the order parameter dynamics is not coupled to these flows. 
For example, if the order parameter represents the concentration difference between two lipid species, it is clear that besides diffusion, 
the order parameter is advected by the lipid flow in the membrane~\cite{Reynwar2008}.

The coupling between the order parameter $\phi(\rv)$ and the hydrodynamic flow $\uv(\rv)$ has been modeled by Camley and Brown~\cite{Camley2012}. 
They found that the advection of the order significantly changes the total drag on the protein. As a result, 
the protein is endowed with a larger effective radius, since the lipids which are closest to the protein are carried along with it. 
Although this effect will affect the value of the drag, this will not change the qualitative behavior, in particular 
the result that the drag should scale with the particle size, in a
 logarithmic way in the SD regime and linearly for large particles.

Recently, another very interesting hydrodynamic model~\cite{Morris2015}-\cite{Daniels2016} has been proposed to account for the experimental 
data on the tension dependence of the mobility of KvAP discussed in this paper and first reported in Ref.~\cite{Quemeneur2014}. 
Assuming a negligible contribution due to the hydrodynamics of the surrounding fluid, the authors of this study 
have exploited a covariant formulation of the membrane hydrodynamics of the inclusion, which has specific 
features due to the Gaussian curvature induced locally by the inclusion. 
In this way, they could explain the experimental data on the mobility of KvAP 
without requiring additional mechanism of internal dissipation.
Interestingly, a key ingredient of their model is the deformability of the inclusion, 
which is particularly important for gated membrane channels \cite{Morris2017}.
We note that the polaron model could account for the deformability of the inclusion, 
and could also be combined with such hydrodynamic description of the membrane.

\section{Concluding perspective}

In this chapter, we have presented some simple views on the way geometrical or mechanical factors influence the mobility of membrane proteins. 
We have shown that one should expect generally a significant dependence of the protein mobility on its local environment.

Naturally, many molecular details affect this local environment, which makes the modeling of the protein mobility a challenging task.  
However, despite the complexity of this problem, there is a simple take-home message. In order to predict the way various mechanical 
factors affect the mobility of membrane proteins, one should not focus on the mechanical factors themselves or on the details of the local environment of the protein, 
but instead on the form of coupling between the membrane and the protein.
We have provided in this chapter a list of various relevant couplings, and  
a detailed study of one coupling, namely the tension dependent coupling 
to the local membrane curvature. 
Clearly, it would be desirable to perform similar experimental studies in order to test more 
couplings and their dependence to other mechanical factors. The importance of these couplings 
goes beyond the specific question of the mobility of membrane proteins, since
they are also relevant to understand the function of many membrane proteins, 
like voltage-gating channels, mechano-sensitive channels, G-protein coupled receptors (GPCR) 
or light-activated rhodopsins \cite{Phillips2009}. 
These couplings also play a key role in the interactions between membrane proteins, and in their 
self-organization into complex dynamical structures, such as membrane clusters. 

At the level of single membrane proteins, other directions for future research, 
should investigate (i) a protein diffusing in a membrane which may be driven out of equilibrium, 
or (ii) a protein which can be activated or can change conformation while diffusing on the membrane. 
For studying the former case (i), various strategies can be used to drive a membrane out of equilibrium by applying external fields on it directly 
or by embedding other proteins in it on which forces can be applied \cite{Lacoste2013}. Measurements of the force dependent 
mobility of a membrane protein could advantageously 
provide information on the way it couples to its environment, which as mentioned above is a crucial aspect of the problem. 
For case (ii), the internal dynamics of the protein can matter for the protein mobility because different internal states can couple differently to the membrane and 
the characteristic time of these internal transitions (typically 10 to 1000 ms) can be much longer than the time of diffusion of the protein 
over a distance equal to its radius (typically < 1ms) \cite{Bouvrais2012}. For many membrane proteins, the transition rates can be tuned by changing the amount 
of ATP, allowing to probe different dynamical regimes for the mobility of the protein. 

In cell membranes, the local environment of proteins is crowded and heterogeneous. 
Membrane protein crowding is of primary importance
to understand the mobility of membrane proteins and 
membrane mediated interactions between different proteins. 
Such effects are not well understood and they need to be investigated more systematically 
both experimentally and theoretically. 

We hope that this chapter can be useful to motivate further experimental and theoretical studies on these questions.

\section*{Acknowledgement}
We would like to thank P. Quemeneur, J. K. Sigurdsson, M. Renner, P. J. Atzberger and P. Bassereau for a previous collaboration, 
which motivated this chapter. In addition, we would like to acknowledge stimulating discussions with 
W. Urbach and M. S. Turner. D L would also like to thank Labex CelTisPhysBio (N ANR-10-LBX-0038) part of IDEX PSL (NANR-10-IDEX-0001-02 PSL)
for financial support.


\end{document}